\documentclass[conference]{IEEEtran}
\usepackage[noadjust]{cite}
\usepackage{amsmath,amssymb,amsfonts}
\usepackage{algorithmic}
\usepackage{graphicx}
\usepackage{textcomp}
\usepackage{siunitx}
\usepackage{xcolor}
\usepackage{url}
\usepackage[hidelinks]{hyperref}
\usepackage [english]{babel}
\usepackage [autostyle, english = american]{csquotes}
\MakeOuterQuote{"}
\def\BibTeX{{\rm B\kern-.05em{\sc i\kern-.025em b}\kern-.08em
    T\kern-.1667em\lower.7ex\hbox{E}\kern-.125emX}}
\begin{document}

\title{Towards Architecting Sustainable MLOps: A Self-Adaptation Approach}

\author{\IEEEauthorblockN{Hiya Bhatt\IEEEauthorrefmark{1}\IEEEauthorrefmark{2} \textsuperscript{\textsection},
Shrikara Arun\IEEEauthorrefmark{2}\textsuperscript{\textsection}, Adyansh Kakran\IEEEauthorrefmark{2} and
Karthik Vaidhyanathan\IEEEauthorrefmark{2}}
\IEEEauthorblockA{\IEEEauthorrefmark{1}Manipal University Jaipur (MUJ)
\IEEEauthorrefmark{2}Software Engineering Research Center,
IIIT Hyderabad\\
hiyabhatt2002@gmail.com, shrikara.a@students.iiit.ac.in, adyansh.kakran@research.iiit.ac.in, \\ karthik.vaidhyanathan@iiit.ac.in
}}

\maketitle
\begingroup\renewcommand\thefootnote{\textsection}
\footnotetext{Equal contribution}
\endgroup

\newcommand{\usecase}{Case Study}
\newcommand{\squeezeup}{\vspace{-2.5mm}}
\newcommand{\supersqueezeup}{\vspace{-4mm}}

\begin{abstract}
In today's dynamic technological landscape, sustainability has emerged as a pivotal concern, especially with respect to architecting Machine Learning enabled Systems (MLS). Many ML models fail in transitioning to production, primarily hindered by uncertainties due to data variations, evolving requirements, and model instabilities. Machine Learning Operations (MLOps) offers a promising solution by enhancing adaptability and technical sustainability in MLS. However, MLOps itself faces challenges related to environmental impact, technical maintenance, and economic concerns. Over the years, self-adaptation has emerged as a potential solution to handle uncertainties. This paper introduces a novel approach employing self-adaptive principles integrated into the MLOps architecture through a MAPE-K loop to bolster MLOps sustainability. By autonomously responding to uncertainties, including data, model dynamics, and environmental variations, our approach aims to address the sustainability concerns of a given MLOps pipeline identified by an architect at design time. Further, we implement the method for a Smart City use case to display the capabilities of our approach.
\end{abstract}

\newcommand{\magenta}[1]{\textcolor{magenta}{#1}}


\section{Introduction}
The concern about the sustainability of software systems has been exacerbated by the emergence of Machine Learning-enabled systems (MLS), which are software systems that incorporate ML models. This is due to their computational complexity and uncertainties arising from their probabilistic nature.
Studies like those by Gartner \cite{gartner2020} show that almost half of the ML models do not successfully transition from prototype to production because they are not sustainable. Sustainability encompasses four dimensions: environmental, technical, social and economical~\cite{beckersustainability}. Environmental concerns, including energy consumption and carbon emissions, are particularly pertinent as ML models often require significant computational resources while training, retraining and deployment. Technical concerns involve ensuring the maintainability \cite{sculleyhidden} and reliability of both the ML models (the core of MLS) and the pipelines (the sequences of data processing and learning steps), which can be challenging given the rapidly evolving nature of ML. Moreover, MLS faces multiple Social concerns like fairness, privacy, explainability, and broader issues like ethics and legislation. Economic concerns are tied to the cost of training, testing, and inference.

\noindent Machine Learning Operations (MLOps) \cite{symeonidis} aims to enhance the technical sustainability of an MLS 
by architecting MLOps pipelines, which include the development, deployment, and maintenance of ML models.
\noindent However, MLOps needs to be enhanced to address the other sustainability concerns. These sustainability concerns for the MLOps pipelines can be identified by an architect at design time and visually represented in a Decision Map (DM)~\cite{Lago}.

\begin{figure}[t!]
\centerline{\includegraphics[width=0.32\textwidth]{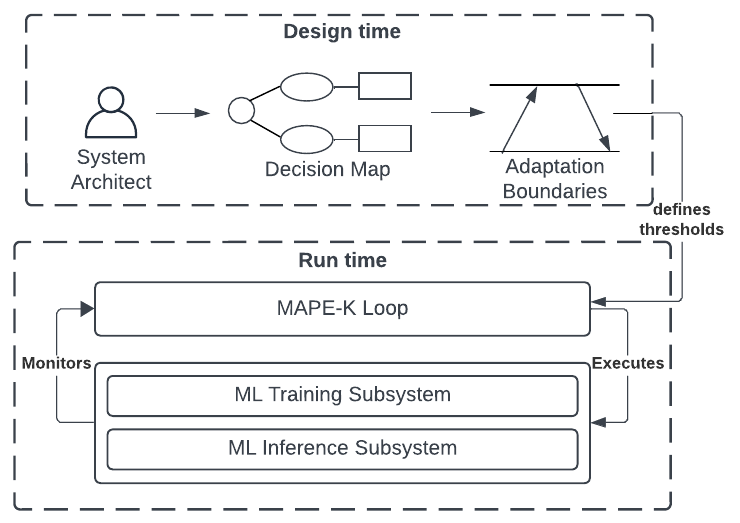}}
\caption{Flow diagram for our approach}
\vspace{-2mm}
\label{flow}
\vspace{-4mm}
\end{figure}

\noindent Although a DM helps capture the sustainability concerns at design time, some of the concerns arise from runtime uncertainties inherent to MLOps pipelines.
These include drifts in data quality, model quality, evolving business needs, and the dynamic environmental context, as put forward by \cite{camara2022uncertainty} \cite{casimiro2022towards}. Over the years, self-adaptation has emerged as a potential solution to handle runtime uncertainties~\cite{weyns2020introduction}. However, not much work has been done on integrating self-adaptation with MLOps to make it sustainable. 

\noindent The research question we pose is \textit{"(How) can self-adaptation be used in MLOps to improve  sustainability?"} To this end, this paper proposes a novel approach that uses self-adaptation in MLOps pipelines through a MAPE-K loop~\cite{kephartvision}  to improve their sustainability.
As in Figure \ref{flow}, the system architect creates a DM at design time, from which adaptation boundaries are realised. This information is stored in the Knowledge base of MAPE-K and enables run time self-adaptation of the MLOps pipelines to enhance sustainability across dimensions. We further demonstrate the practicality of our approach using a Smart City case study to predict Air Quality.

\section{Related Work}
There has been significant effort related to engineering MLS through MLOps, including \cite{shankar2022operationalizing, symeonidis, amershisoftware}. Traditional approaches do not address sustainability through self-adaptation. The closest work we find is \cite{bayram2022drift} describing a self-adaptive approach to handle drift in an industrial process. Our approach differs since it deals not only with model drift but also with other sustainability aspects.
What little work on self-adaptive MLOps that exists fails to address the concern of sustainability, across such dimensions mentioned in this paper, such as  \cite{casimiro2021self}, who apply a MAPE-K loop, and deal entirely with the method and effects of self-adaptation in MLS. Whereas \cite{Tamburri} discusses MLOps and AI Software sustainability, it is not through the lens of runtime self-adaptation. The concept of runtime goal management is described by \cite{Gerostathopoulos} by expressing adaptation intent as a sustainability goal. They propose an approach that uses decision maps to make sustainability-driven decisions for adaptation in a systematic way but does not deal with MLS and the added uncertainties they add due to their probabilistic nature. Switching the lens to the field of software sustainability, \cite{Lago} describes decision maps for software sustainability and categorizes concerns into social, technical, environmental, and economic; and discusses immediate, enabling, and systemic impacts. This, too, does not deal with MLS. Differently from the above works, we propose an approach that aims to enhance the sustainability of MLOps pipelines and thereby MLS by taking into consideration design time goals through runtime self-adaptation powered by MAPE-K loop.

\section{\usecase{}}\label{usecase}
The Smart City Living Lab of IIITH\footnote{\url{https://smartcitylivinglab.iiit.ac.in} (Last accessed 10 Feb, 2024)} is a research platform and test bed for smart city applications comprising more than 300 IoT nodes spanning various domains such as air quality, water quality and quantity, solar power monitoring, home automation, etc. Due to increasing vehicular traffic and pollution, air quality monitoring (involves monitoring of Air Quality Index, AQI) has emerged as one of the key domains as this can contribute to the establishment of environmental regulations in urban India.
As part of this domain, there are 10 outdoor and 5 indoor sensor nodes deployed in the campus which collect air quality information such as PM2.5, PM10 levels, temperature and humidity at a target frequency of once every second. 
These data are then used by ML pipelines to forecast AQI~\cite{nileshiot}. However, these pipelines do not take into consideration the sustainability concerns arising from model and data drifts (technical), increased cost and energy consumption due to frequent retraining (economical and environmental). 
To this end, we envision architecting a self-adaptive MLOps pipeline to enhance  sustainability. We use this case study in the remainder of this paper to explain our approach and the results obtained through our preliminary evaluations. 

\section{Defining Our Approach}
Our architecture utilizes the MAPE-K framework for self-adaptation to respond to detected uncertainties.
The three components of our approach are: the Managed System, which is the MLS, the Managing System, which is responsible for monitoring the system and its environment, detecting uncertainties at runtime, planning and executing adaptations; and the Decision Map, which captures the sustainability concerns of the system at design time.

\begin{figure}[htbp]
\centerline{\includegraphics[width=0.33\textwidth]{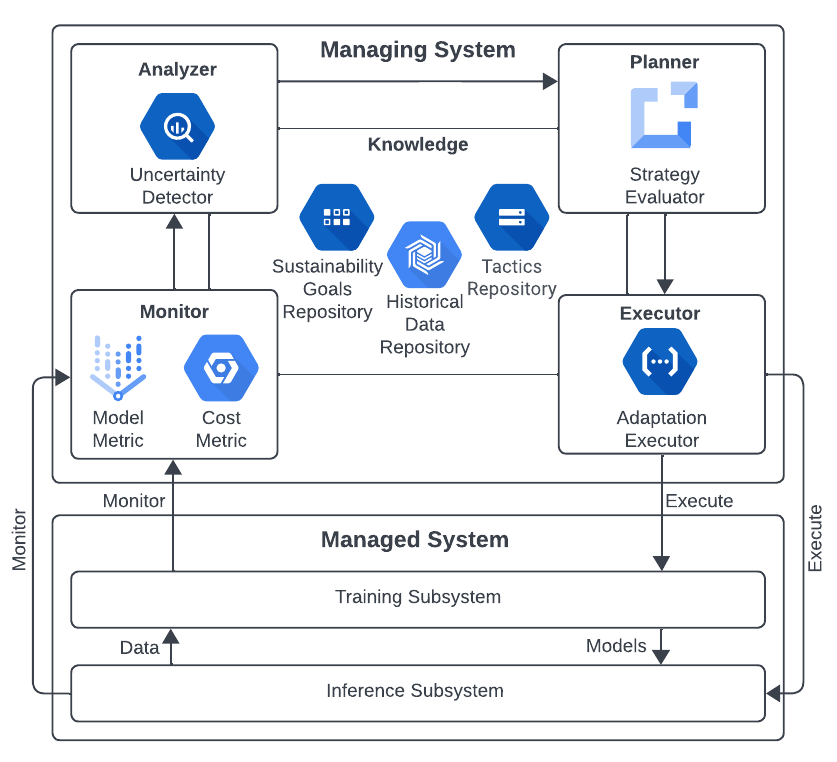}}
\caption{Approach}
\label{fig1}
\vspace{-5.5mm}
\end{figure}

\subsection{Managed System}
\noindent The Managed System encompasses the entirety of the MLOps pipeline, consisting of two components, as in \cite{ubermichelangelo}:

\noindent\textit{1) Training Subsystem} is responsible for developing and refining models based on available data. It handles model and data versioning to track changes and improvements. 

\noindent\textit{2) Inference Subsystem} serves trained models obtained from the training subsystem to deliver online/batch predictions.

\subsection{Decision Map}
\begin{figure}[htbp]
\vspace{-2mm}
\centerline{\includegraphics[width=0.8\columnwidth]{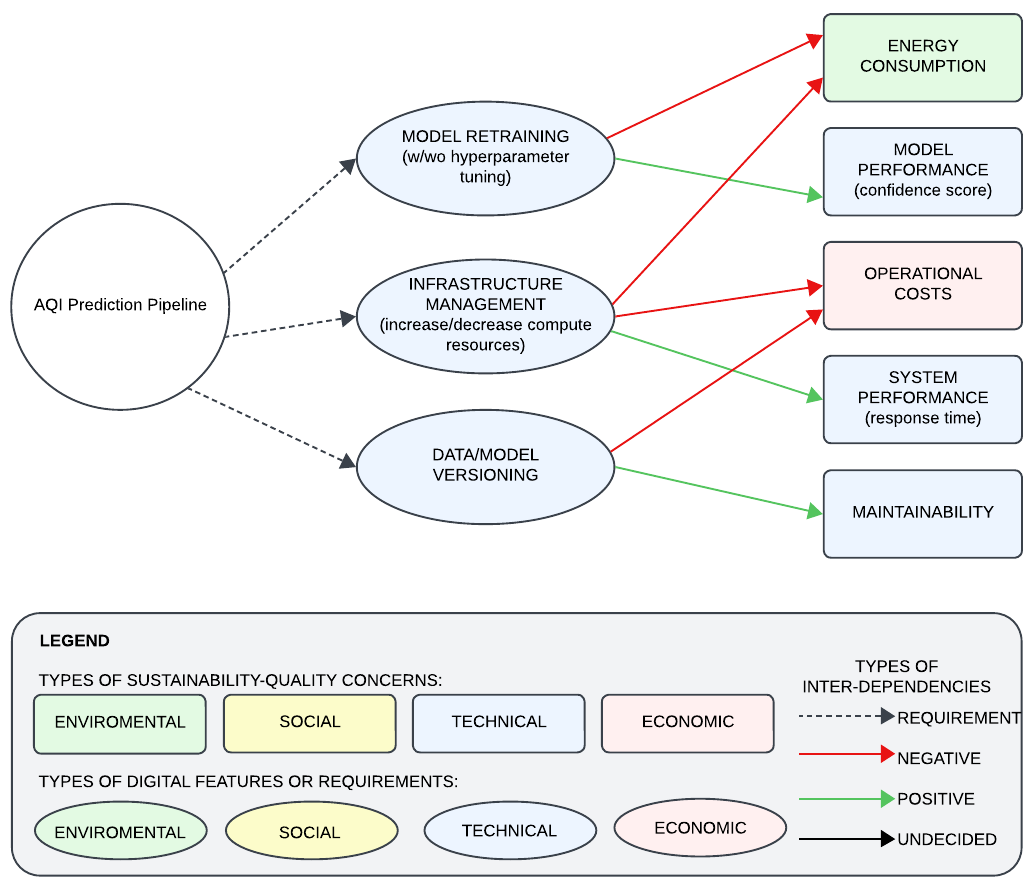}}
\caption{Decision Map for AQI Prediction Pipeline}
\label{DM}
\vspace{-2.3mm}
\end{figure}

\noindent A Decision Map (DM), as shown in Figure \ref{DM}, is a visual representation of sustainability concerns across the Social, Environmental, Technical and Economic dimensions~\cite{Lago}.  
A system architect decides the sustainability goals for each concern during design time, setting the stage for runtime adaptation. The architect also defines and modifies adaption boundaries which dictate the acceptable quality of the system across sustainability dimensions \cite{Gerostathopoulos}. 
For instance, $(rt_{max}, rt_{min})$, where $rt_{min}$ and $rt_{max}$ denote the minimum and maximum allowed response time of the system.

\begin{table*}[htbp]
\centering
\caption{Self-Adaptation Uncertainties, Tactics and Strategies for Sustainable MLOps \\(those in bold have been implemented in the \usecase{})}

\begin{tabular}{|>{\raggedright\arraybackslash}p{1.7cm}|>{\raggedright\arraybackslash}p{1.6cm}|>{\raggedright\arraybackslash}p{1.4cm}|>{\raggedright\arraybackslash}p{2cm}|>{\raggedright\arraybackslash}p{1.8cm}|>{\raggedright\arraybackslash}p{2.2cm}|>{\raggedright\arraybackslash}p{4.2cm}|} \hline 

 \textbf{Uncertainty} & \textbf{Concern}&\multicolumn{3}{|c|}{\textbf{Impact}} & \textbf{Tactics} & \textbf{Strategy} \\  \cline{3-5}
 & & \textbf{Immediate} & \textbf{Enabling} & \textbf{Systemic} & & \\ \hline 

 1. \textbf{Model Drift}& \textbf{Technical}&Reduced quality of predictions& Reduced user trust& Reduced social benefits& \textbf{1. Model switch
\newline\newline\_\_\_\_\_\_\_\_\_\_\_\_\_\_\_\_\_\_\_\_\_
2. Retrain}& \textbf{Switch to a model with better performance but higher energy use}
\_\_\_\_\_\_\_\_\_\_\_\_\_\_\_\_\_\_\_\_\_\_\_\_\_\_\_\_\_\_
\newline
a. Conduct incremental learning \newline b. Conduct transfer learning \newline c. \textbf{Conduct complete retraining (with/without      different hyperparameters)}\\ \hline 

 2.\textbf{ High energy consumption}& \textbf{Environmental}&Increased costs& Increased environmental impact& Reduced economic competitiveness& \textbf{Model switch}& \textbf{Switch to a model with lower energy consumption}\\ \hline 

 3. Future goals changes& Technical&Increased uncertainty: system's purpose& Difficulty in maintaining the system& Reduced public trust in AI& Update model objectives& Change thresholds for: \newline a. Business metrics \newline b. Cost metrics \newline c. Model metrics\\ \hline 

 4. Rise in cost& Economic&Increased costs& Difficulty in deploying and maintaining the system
& Reduced ability to compete in the market& Optimize the system to reduce costs & a. Reduce the amount of resources used by the system \newline b. Use more cost-effective resources \\ \hline 
\end{tabular}
\label{tab1}

\vspace{-4mm}
\end{table*}

\noindent The DM in Figure \ref{DM} describes the sustainability concerns for our \usecase{},
a) \textit{Model Retraining}: 
We would need to retrain our models with the latest data to account for shifts in distribution of AQI caused due to seasons, increased emissions and weather. 
b) \textit{Infrastructure Management}: The hardware infrastructure used to train the model and conduct inference can affect response time, cost and energy utilization.
c) \textit{Data/Model Versioning}: Improves maintainability and reproducibility, but can increase operational costs.

\noindent Considering these concerns, we identify some uncertainties from \cite{camara2022uncertainty} that affect them, as well as their \textit{immediate, enabling }and\textit{ systemic impacts} in Table \ref{tab1}. Table \ref{tab1} also contains the tactics to mitigate identified uncertainties and the strategies that can be employed to carry out the tactics, more of which can be created by the architect.

\subsection{{Managing System}}

\noindent\textbf{Knowledge:}
The Knowledge base, as in Figure \ref{fig1} is divided into three subsections: 
1) {\em Sustainability Goals Repository} stores the sustainability goals defined by the system architect during the design phase, and the acceptable threshold for metrics derived from the DM which are defined by the adaptation boundaries. 
2) {\em Historical Data Repository} stores the version history of models and data. 
3) {\em Tactics Repository} stores mitigation tactics and the corresponding strategies.

\noindent 
\textbf{Monitor: }The Monitor component, as in Figure \ref{fig1} continuously collects data about the managed system's state, including cost metrics and model metrics. 
The {\em Cost Metrics} include the cost of compute resources, data storage, model training, and inference. The {\em Model Metrics} evaluate the model's performance, including its confidence/accuracy and energy consumption.

\noindent \textbf{Analyzer:} 
The core of the Analyzer component is the {\em Uncertainty Detector} (refer Figure \ref{fig1}), which analyzes the metrics collected by the Monitor for anomalies or trends that might signal uncertainties.
These uncertainties, as defined in Table \ref{tab1}, are flagged based on thresholds in the \textit{Sustainability Goals Repository}. 
The {\em Uncertainty Detector} assesses the potential impact of these uncertainties on the managed system to determine if an adaptation is necessary. These adaptation boundaries delineate the acceptable quality of the system~\cite{Gerostathopoulos}. These boundaries are not fixed and can be adjusted based on factors such as the system's current state, changing requirements, and environmental conditions. This makes it more effective at detecting anomalies and other changes that may indicate uncertainty since our definition of uncertainty, too, may evolve. For instance, 
assume that in our case study (refer Section \ref{usecase}), the energy consumed for performing the forecasts, $E$, violates the sustainability goals defined for energy consumption such that $E \leq E_{min}$ or $E \geq E_{max}$, 
then we can say that the quality of the system has deteriorated, and the Planner will be triggered to decide the adaption strategy. 
\\
\noindent \textbf{Planner:} When the Analyzer detects an uncertainty, the Planner component determines the optimal strategy to address it using the {\em Strategy Evaluator}. It decides on the tactic and strategy the Managing system should employ by mapping the detected uncertainties to their respective tactic and strategy.
For instance, in our case study, if the trigger for the planner is due to violations of energy, then the planner may switch the model (refer Table \ref{tab1}).
Additionally, more sophisticated planner strategies can incorporate Reinforcement Learning (RL) \cite{Metzger2022}, model checkers \cite{model-checker} or clustering \cite{kulkarni2023towards} methods in combination with the aforementioned methods. \\
\noindent \textbf{Executor: } 
The {\em Adaptation Executor} in the Executor implements the strategy selected by the Planner component. It interacts with both the training and the inference subsystem. 
For instance, it can trigger a (re)training job in the training subsystem or can switch to a different model in the inference subsystem based on the selected strategy.

\section{Preliminary Results}
We implemented
our approach 
(by addressing the model drift \& high energy consumption uncertainties which affect the technical \& environmental sustainability concerns respectively)
on the AQI forecasting pipeline (Section \ref{usecase}). 
We utilize the DM in Figure \ref{DM} to realize adaptation boundaries, which are mentioned in the GitHub repository\footnote{\url{https://github.com/sa4s-serc/sustainableMLOps}}. 
The forecasting pipeline consists of two different models,
one which uses Linear Regression (lightweight model with low response time \& prediction quality) \& one which makes use of a Long Short Term Memory network (LSTM) (heavier model with high prediction quality \& response time.)

\begin{figure}[htbp]
\vspace{-2mm}
\includegraphics[width=\columnwidth]{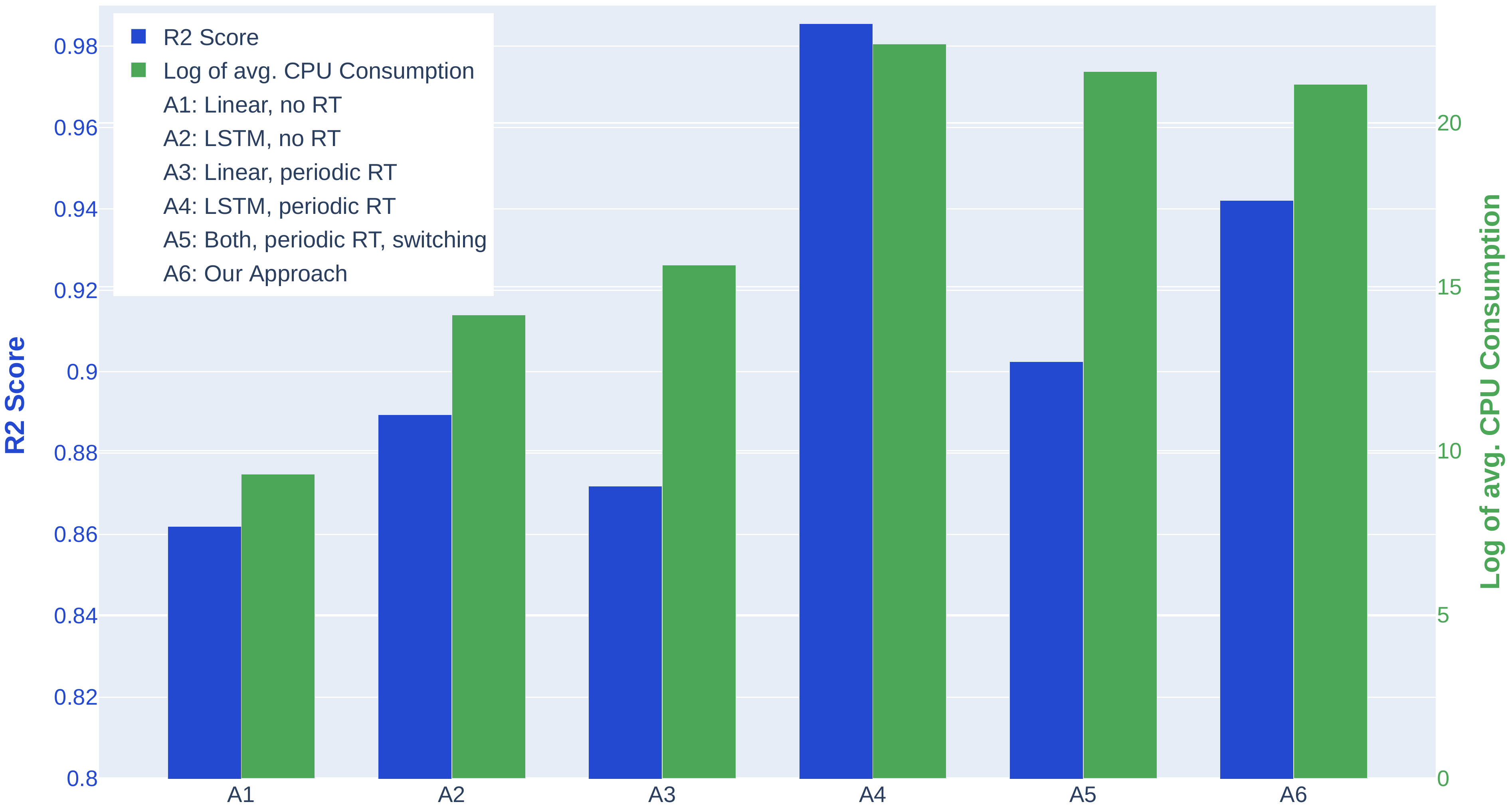}
\caption{$R^2$ Score \& Log of Average CPU Consumption over $10s$ ($\mu J$). RT=retraining} 
\label{Result}
\vspace{-2mm}
\end{figure}
\noindent We further evaluated our approach against five baseline approaches: two that use Linear Regression \& LSTM, each without retraining (RT) (A1 \& A2), two that use periodic RT \cite{Nazirarchitecting} (A3 \& A4), and one that uses both models with periodic RT \& switching (A5). We switch between the models based on CPU consumption. As shown in Figure \ref{Result}, our approach (A6) strikes a balance between performance, measured by $R^2$ score and average CPU consumption over the past $10$s, measured in $\mu J$. Unlike A3, A4 \& A5, we retrain the models only when model drift is detected. We calculate the model drift using KL divergence, which quantifies the disparity between the distribution of training data and real-time encountered data. While using only the LSTM with periodic retraining (A3) offers the best $R^2$ score, it consumes significantly more energy than other approaches. 
Our approach, as compared to periodically retraining both models and switching between them (A5), improves $R^2$ score from $0.90$ to $0.94$ and reduces average CPU consumption by $32\%$.

\section{Conclusion and Future Work}
In this paper, we propose an architecture for a sustainable MLOps pipeline by utilizing self-adaptation through a MAPE-K loop. During design time, a system architect identifies adaptation boundaries through the creation of a decision map to allow for runtime self-adaptation. Preliminary evaluations in our case study show that our approach strikes a balance between performance and CPU consumption. We believe that a self-adaptive MLOps architecture can pave the way to increase sustainability of MLOps pipelines. Future work includes evaluating the generalizability of our approach to different domains and identifying more uncertainties and alleviation tactics, especially in the Social dimension. While we focus on supervised ML tasks, work also needs to be done to address unsupervised and reinforcement learning tasks. The rapid landscape of generative AI also presents promising avenue for further research.

\bibliographystyle{IEEEtran}
\bibliography{references}

\end{document}